\documentclass[12pt]{article}

\usepackage{amssymb,amsmath}

\newcommand{\be}{\begin{equation}}
\newcommand{\ee}{\end{equation}}

\newcommand{\dlt}{\delta}

\newcommand{\bt}{\beta}
\newcommand{\vp}{\varphi}
\newcommand{\ep}{\varepsilon}
\newcommand{\al}{\alpha}
\newcommand{\ra}{\rightarrow}
\newcommand{\sgm}{\sigma}

\newcommand{\om}{\omega}

\newcommand{\cH}{{\cal H}}

\newcommand{\cA}{{\cal A}}

\newcommand{\rgl}{\rangle}
\newcommand{\lgl}{\langle}

\begin{document}

\begin{center}
{\Large{\bf
Decoherence and equilibration under nondestructive measurements} \\ [5mm]
V.I. Yukalov} \\ [3mm]

{\it Bogolubov Laboratory of Theoretical Physics, \\
          Joint Institute for Nuclear Research, Dubna 141980, Russia}
\end{center}

\vskip 3cm

\begin{abstract}                                    
The evolution of observable quantities of finite quantum systems is 
analyzed when the latter are subject to nondestructive measurements. 
The type and number of measurements characterize the level of decoherence
produced in the system. A finite number of instantaneous measurements 
leads to only a partial decoherence. But infinite number of such 
measurements yields complete decoherence and equilibration. Continuous 
measurements result in partial decoherence in finite time, but produce 
complete decoherence and equilibration as time tends to infinity. 
Resulting equilibrium states are characterized by representative 
statistical ensembles that, generally, retain information on initial 
conditions. Any system, to be observable, necessarily requires the 
presence of measurements, whose large number leads to the system 
equilibration and decoherence. 
\end{abstract}

\vskip 2cm

{\bf PACS}: 03.65.Yz, 03.65.Ta	

\vskip 1cm

{\bf Keywords}: Decoherence; Equilibration; Quasi-open systems; 
Nondestructive measurements

\vskip 1cm
{\bf E-mail address}: yukalov@theor.jinr.ru

\newpage

\section{Introduction}

Decoherence and equilibration in quantum systems are the problems that has 
been studied in numerous papers. An extensive list of related literature can 
be found in the recent review articles [1-13]. Nowadays, this problem has 
gained much interest with regard to finite quantum systems. Such finite systems 
are now intensively studied both theoretically and experimentally because of 
their role in a variety of applications ranging from quantum electronics to 
quantum information storage, processing, and computing [14-17]. Equilibration 
and decoherence from a strongly nonequilibrium initial state have been studied, 
e.g., for such quantum systems as spin assemblies [7,18-21], trapped atoms 
[6,13,22-27], and quantum dots [28]. 

One, generally, distinguishes two kinds of system dynamics: one is when the 
finite quantum system is isolated and another when it is connected to some
environment. It is known that if the system is coupled to a sufficiently 
large equilibrium environment, it equilibrates and decoheres due to its 
interaction with surrounding [2,4,29-32]. But if the system is isolated, 
its motion is quasi-periodic and, thus, the system cannot equilibrate in the 
strict sense, displaying instead the Poincar\'e recurrences [33]. However, 
an isolated system can equilibrate on average, relaxing to a quasi-equilibrium 
state defined by an ergodic average [34] and staying close to it most of the 
time [5]. Dynamics of isolated quantum systems depends on their closeness
to integrability, revealing pre-equilibration effects [18,19,35,36] and 
displaying essential dependence on the way of their preparation, especially 
on the presence of defects [37,38].

In the present paper, we consider the intermediate case, when a finite quantum 
system is quasi-isolated, being almost isolated, except the action of 
nondestructive measurements, so that they do not destroy the system properties. 
We analyze the system decoherence under such nondestructive measurements. 
The main difference of the present consideration from the previous works is in
the following.

   (i) The studied quantum system interacts not with an equilibrium bath, but 
with a nonequilibrium measuring device. 

   (ii) The interaction part of the Hamiltonian is time-dependent, while
a bath is usually described by a time-independent Hamiltonian. This essentially 
complicates mathematics and results in rather different consequences. The 
evolution operator now is not the standard exponential form $\exp(-Ht)$. Solving
the evolution equation for this operator requires now to invoke the 
Lappo-Danilevsky theory.

   (iii) In the case of measurements, decoherence is not necessarily complete,
as it would be for the bath, but the level of decoherence depends on the type 
and number of measurements.

Specifically, nondestructive measurements are analyzed, but not arbitrary 
external perturbations of the system. In that sense, the consideration is limited 
by exactly this type of measurements. Under this restriction, the overall  
treatment can be done for quite general conditions: the nature of the system can 
be arbitrary; it is not required that its spectrum be nondegenerate; the system 
states can be either pure or mixed; it is not required that the initial states be 
of the product type; no time averaging is needed; interactions of the system with 
the measuring device can be of arbitrary strength, but not necessarily weak; the 
measuring device can also be of rather arbitrary nature, provided it is 
nondestructive. 
 
As is stressed above, the main point is the consideration of the time-dependent
interaction of the system with an external device, which makes the principal 
difference from the standardly treated case of an equilibrium bath. Time-dependent 
interactions are typical of measurement procedures, when a measuring device is 
switched on and off. This is why it is possible to interpret such nonequilibrium 
interactions as measurements. Throughout the paper, the word "measurement" is used
as a conditional term describing time-dependent influence on the system of an 
external nonequilibrium source. In addition, two limiting cases of the source
influence, instantaneous and continuous, which will be analyzed in the paper, have 
the forms that are commonly associated with measurements [39-41]. Therefore the 
interpretation of the time-dependent source influence as measurement seems to be 
justified. This, however, is not compulsory and one can treat the word "measurement"
just as a brief term for the considered time-dependent interactions possessing 
several of the properties typical of measurement process.

\section{Quantum system under nondestructive measurements}

Let the quantum system of interest be described by a Hamiltonian $H_A$ acting on 
a Hilbert space $\cH_A$. The system is subject to a measurement procedure. The 
measuring device is characterized by a Hamiltonian $H_B$ acting on a Hilbert 
space $\cH_B$. The total Hamiltonian is the sum
\be
\label{1}
H_{AB} = H_A + H_B + H_{int} \; ,
\ee
where $H_{int}$ is the term describing the interaction between the system and the
measuring device. Hamiltonian (1) is defined on the Hilbert space
\be
\label{2}
 \cH_{AB} = \cH_A \bigotimes \cH_B \;  .
\ee
Strictly speaking, the Hamiltonian (1) should be written as
\be
\label{3}
H_{AB} = H_A \bigotimes \hat 1_A + \hat 1_A \bigotimes H_B + 
H_{int} \;   ,
\ee
with $\hat{1}_A$ and $\hat{1}_B$ being the unit operators on the corresponding 
spaces. However, it is commonly accepted to omit the unit operators, writing,
for simplicity, the Hamiltonian in form (1).  

The measurement is called {\it nondestructive} when it does not disturb the 
system, in the sense that the system Hamiltonian is conserved,
\be
\label{4}
 [ H_A , \; H_{AB} ] = 0 \; .
\ee
Conversely, the studied system does not destroy the measuring device, so that
its Hamiltonian is also conserved,
\be
\label{5}
 [ H_B , \; H_{AB} ] = 0 \;   .
\ee
As a consequence of this definition, one has
\be
\label{6}
 [ H_A , \; H_{int} ] = [ H_B , \; H_{int} ] = 0 \; , \qquad
[ H_{int}, \; H_{AB} ] = 0 \;  .
\ee
The measuring procedure, enjoying these properties, can also be termed 
minimally disturbing measurement [42-44] or nondemolition measurement [45-48].    

The system Hamiltonian defines a complete orthonormal basis $\{|n\rgl\}$ by
the eigenproblem
\be
\label{7}
 H_A | n \rgl = E_n | n \rgl \;  .
\ee
Respectively, the device Hamiltonian, by the eigenproblem 
\be
\label{8}
H_B | k \rgl = \bt_k | k \rgl \;   ,
\ee
defines a complete orthonormal basis $\{|k\rgl\}$.

Because of properties (4) and (5), the total Hamiltonian (1) possesses the 
eigenfunctions $|nk\rgl\equiv |n\rgl\bigotimes |k\rgl$. A measurement procedure 
is a nonequilibrium process, which leads to a complication coming from the fact 
that the interaction Hamiltonian $H_{int}=H_{int}(t)$, generally, depends on 
time. Therefore the eigenvalues of the total Hamiltonian also depend on time, 
being given by the eigenproblem
\be
\label{9}
 H_{AB}(t) | nk \rgl = [ E_n + \ep_{nk}(t) ] \; | \; nk \rgl \;  .
\ee

The temporal evolution of the total statistical operator is prescribed by
the rule
\be
\label{10}
 \hat\rho_{AB}(t) = \hat U_{AB}(t) \hat\rho_{AB}(0) 
\hat U_{AB}^+(t) \;  ,
\ee
where the evolution operator is a unitary operator satisfying the 
Schr\"{o}dinger equation
\be
\label{11}
 i\; \frac{d}{dt} \; \hat U_{AB}(t) = 
H_{AB} \hat U_{AB}(t) \;  .
\ee

The total Hamiltonian $H_{AB} = H_{AB}(t)$ depends on time through the 
interaction term $H_{int}(t)$. This does not allow us to represent the 
evolution operator in a simple exponential form as it is usually accepted 
in the case of an equilibrium bath. However, we may notice that 
$$
\left \lgl mk \left | H_{AB}(t) \int_0^t H_{AB}(t') \; dt' 
\right | np \right \rgl =
$$
\be
\label{12}
= \dlt_{mn} \dlt_{kp} \; [\; E_n + \ep_{nk}(t)\; ]
\int_0^t [\; E_n + \ep_{nk}(t') \;] \; dt' \;   .
\ee
Consequently, the Lappo-Danilevsky condition [49] 
\be
\label{13}
\left [ H_{AB}(t) , \; \int_0^t H_{AB}(t') \; dt' \right ] = 0   
\ee
holds true, at least in the weak sense. Owing to this condition (13), one 
can represent the solution of Eq. (11) as
\be
\label{14}
 \hat U_{AB}(t) = \exp \left \{ - i \int_0^t H_{AB}(t') \; dt' 
\right \} \;  .
\ee
The possibility of representing the evolution operator in form (14) is 
due to the Lappo-Danilevsky condition (13) that becomes valid when the 
measurement procedure is nondestructive in the sense of conditions (4) 
and (5).

Thus the total composite system is characterized by the statistical ensemble 
$\{\cH_{AB}, \hat{\rho}_{AB}(t)\}$, with the statistical operator 
satisfying the normalization condition
\be
\label{15}
{\rm Tr}_{AB} \hat\rho_{AB}(t) = 1 \; ,
\ee
where the trace is over the total space (2).

\section{Temporal evolution of observable quantities}

What one is interested in any physical problem is the behavior of the system 
observable quantities that are represented by self-adjoint operators $\hat A$ 
composing the algebra of local observables $\cA \equiv \{ \hat A \}$ 
defined on the system space $\cH_A$. The measurable observables are given by 
the averages
\be
\label{16}
\lgl \hat A(t) \rgl \equiv {\rm Tr}_{AB} \hat\rho_{AB}(t) \hat A \;  ,
\ee
in which the trace is over $\cH_{AB}$. But, since $\hat{A}$ is given on 
$\cH_A$, the latter average (16) reduces to
\be
\label{17}
 \lgl \hat A(t) \rgl = {\rm Tr}_{A} \hat\rho_{A}(t) \hat A \; ,
\ee
with the partial statistical operator
\be
\label{18}
 \hat\rho_{A}(t) \equiv {\rm Tr}_{B} \hat\rho_{AB}(t) \; ,
\ee
in which the degrees of freedom of $\cH_B$ are traced out. 

Passing to a matrix representation transforms average (17) to
\be
\label{19}
 \lgl \hat A(t) \rgl = \sum_{mn} \rho_{mn}^A(t) A_{nm} \; ,
\ee
with $A_{mn} \equiv <m|\hat{A}|n>$ and a density matrix
\be
\label{20}
\rho_{mn}^A(t) \equiv \lgl m | \hat\rho_A(t) | n \rgl \; .
\ee
The latter can be written as
\be
\label{21}
 \rho_{mn}^A(t) = \sum_k \rho_{mnk}^{AB}(t) \;  ,
\ee
where
\be
\label{22}
 \rho_{mnk}^{AB}(t) \equiv \lgl mk | \hat\rho_{AB}(t) | nk \rgl \; .
\ee
 
In general, any basis could be used for such a matrix representation [50].
But, as far as the system observables are of interest, it is convenient 
to employ the basis $\{|nk\rgl\}$ composed of the eigenvectors $|n\rgl$ of the 
system Hamiltonian $H_A$ and eigenvectors $|k\rgl$ of $H_B$. Then, because of
Eqs. (9) and (14), one has
$$
 \hat U_{AB}(t) | nk \rgl =
\exp \left \{ - iE_n t - i \int_0^t \ep_{nk}(t') \; dt' 
\right \} | \; nk \; \rgl \; .
$$
This, for matrix (22), gives   
\be
\label{23}
 \rho_{mnk}^{AB}(t) = \rho_{mnk}^{AB}(0) \exp \left \{ 
- i\om_{mn} t - i \int_0^t \ep_{mnk}(t') \; dt' \right \} \;  ,
\ee
with the initial condition
\be
\label{24}
 \rho_{mnk}^{AB}(0) \equiv \lgl mk | \hat\rho_{AB}(0) | nk \rgl \;  ,
\ee
transition frequencies
\be
\label{25}
 \om_{mn} \equiv E_m - E_n \;  ,
\ee
and the notation
\be
\label{26}
  \ep_{mnk}(t) \equiv \ep_{mk}(t) - \ep_{nk}(t) \; .
\ee

By definitions (25) and (26), one has
\be
\label{27}
\om_{nn} = 0 \; , \qquad \ep_{nnk}(t) = 0 \; ,
\ee
because of which the diagonal element
\be
\label{28}
\rho_{nn}^A(t) = \sum_k \rho_{nnk}^{AB}(0) \equiv \rho_{nn}
\ee
does not depend on time. Because of normalization (15), we have 
\be
\label{29}
 \sum_n \rho_{nn}^A(t) = \sum_n \rho_{nn} = 1 \;  .
\ee

Separating in sum (19) the terms with $m = n$ and $m \neq n$ yields
\be
\label{30}
 \lgl \hat A(t) \rgl = \sum_n \rho_{nn} A_{nn} +
\sum_{m\neq n} \rho_{mn}^A(t) A_{nm} \;  .
\ee
 
As an initial expression $\hat{\rho}_{AB}(0)$ one usually takes a disentangled
factor product of $\hat{\rho}_A(0)$ and $\hat{\rho}_B(0)$. Here we do not assume
this simplification, but keep the general form of $\hat{\rho}_{AB}(0)$. Whether 
the latter is entangled or not does not play a principal role for what follows.

\section{Measurement procedure of several measurements}

In general, a measurement procedure can include multiple acts of measurement.
Let there be $M$ such measurement acts, so that the action of the measuring 
device be represented as the sum 
\be
\label{31}
  H_{int}(t) = \sum_{j=1}^M f_j(t) \hat X_j \; ,
\ee
in which $f_j(t)$ is a real function and $\hat{X}_j$ is a self-adjoint operator 
on $\mathcal{H}_{AB}$.

According to the definition of nondestructive measurements in Eqs. (4) and (5) 
and its consequence (6), we have the commutators
\be
\label{32}
 [ H_A \; \hat X_j ] = [ H_B , \; \hat X_j ] = 0 \; .
\ee
Therefore the eigenproblem for $\hat{X}_j$ reads as
\be
\label{33}
\hat X_j | nk \rgl = \xi_{jnk} | nk \rgl \;   ,
\ee
with a real eigenvalue $\xi_{jnk}$. As a result, the eigenproblem for the 
interaction operator (31) takes the form
\be
\label{34}
 H_{int}(t) | nk \rgl = \al_{nk}(t) | nk \rgl \;  ,
\ee
with the eigenvalue
\be
\label{35}
\al_{nk}(t) = \sum_{j=1}^M \xi_{jnk} f_j(t) \;   .
\ee

In view of Eqs.(8) and (9), we get
$$
 \ep_{nk}(t) = \al_{nk}(t) + \bt_k \;  ,
$$
which leads to
$$
\ep_{mk}(t) - \ep_{nk}(t) = \al_{mk}(t) - \al_{nk}(t) \;   .
$$
Then Eq. (26) gives
\be
\label{36}
\ep_{mnk}(t) = \sum_{j=1}^M x_{jmnk} f_j(t) \;   ,
\ee
where
\be
\label{37}
 x_{jmnk} \equiv \xi_{jmk} - \xi_{jnk} \;  .
\ee
  
Introducing the notation
\be
\label{38}
R_{mn}(t) \equiv \sum_k \rho_{mnk}^{AB}(0) \exp \left \{ - i
\sum_{j=1}^M x_{jmnk} \vp_j(t) \right \} \;  ,
\ee
with
\be
\label{39}
 \vp_j(t) \equiv \int_0^t f_j(t') \; dt' \;  ,
\ee
for matrix (21), we obtain
\be
\label{40}
 \rho_{mn}^A(t) = R_{mn}(t) \exp ( - i \om_{mn} t) \;  .
\ee

\section{Evolution after last measurement}

Function (39) characterizes the integral impact of the $j$-th measurement 
during the period of time $[0,t]$. Suppose that after the last $M$-th 
measurement, occurring at time $t_M$, the integral impact (39) becomes 
\be
\label{41}
 \vp_j(t) = \vp(t) \qquad ( t > t_M ) \;  .
\ee
As is shown below, property (41) is valid for different types of measurements,
including discrete measurements, whose action is equivalent to instantaneous 
kicking [51,52], as well as in the opposite case of continuous permanent 
measurements, acting uniformly in time [41,47].   

Under condition (41), Eq. (38) reduces to 
\be
\label{42}
 R_{mn}(t) = \sum_k \rho_{mnk}^{AB}(0) 
\exp \{ -ix_{mnk} M \vp(t) \}\;  ,
\ee
where
\be
\label{43}
x_{mnk} \equiv \frac{1}{M} \; \sum_{j=1}^M x_{jmnk} \; .
\ee

Let us introduce the density function 
\be
\label{44}
g_{mn}(x) \equiv \sum_k \rho_{mnk}^{AB}(0) \dlt(x-x_{mnk} ) \; .
\ee
This function incorporates the properties of the measuring device, which affect 
the measured quantity. Therefore it can be called the {\it effect density} [39].
Definition (44) makes it possible to rewrite Eq. (42) as the integral transformation
\be
\label{45}
 R_{mn}(t) = \int g_{mn}(x) \exp\{ - i x M \vp(t) \}\; dx \;  .
\ee

The effect density (44) is normalized as
\be
\label{46}
 \int g_{mn}(x) \; dx = \sum_k \rho_{mnk}^{AB}(0) \; .
\ee
Invoking the definition
\be
\label{47}
\rho_{mn} \equiv \rho_{mn}^A(0) = \sum_k \rho_{mnk}^{AB}(0)
\ee
reduces normalization (46) to the form
\be
\label{48}
  \int g_{mn}(x) \; dx = \rho_{mn} \;  .
\ee
Therefore the effect density can be represented as
\be
\label{49}
 g_{mn}(x) = \rho_{mn} p_{mn}(x) \;  ,
\ee
where the distribution $p_{mn}(x)$ is normalized to one,
\be
\label{50}
 \int p_{mn}(x) \; dx = 1 \; .
\ee

Thus for Eq.(45), we come to the expression
\be
\label{51}
R_{mn}(t) = \rho_{mn} D_{mn}(t) \;   ,
\ee
with the {\it decoherence factor}
\be
\label{52}
 D_{mn}(t) \equiv \int p_{mn}(x) \exp\{ - i x M \vp(t) 
\} \; dx \; .
\ee
Respectively, matrix (40) reads as
\be
\label{53}
\rho_{mn}^A(t) = \rho_{mn}(t) D_{mn}(t) \;  ,
\ee
where
\be
\label{54}
 \rho_{mn}(t) \equiv \rho_{mn} \exp(-i\om_{mn} t ) \;  .
\ee

As a result, for the evolution of observable quantities (30), we have
\be
\label{55}
\lgl \hat A(t) \rgl = \sum_n \rho_{nn} A_{nn} +
\sum_{m\neq n} \rho_{mn}(t) A_{nm} D_{mn} (t) \;  .
\ee
This formula describes the evolution of observables after the system has 
been subject to a measurement consisting of a series of nondestructive 
measurement acts.

\section{Decoherence caused by nondestructive measurements}

The second term in Eq. (55) is due to interference effects typical of 
coherent quantum systems. Measurement procedure destroys coherence. In 
order to derive an explicit expression for the decoherence factor (52) 
one has to model the distribution function $p_{mn}(x)$ and to define 
the type of the measurement procedure characterized by function $\vp(t)$. 
The measuring device is a macroscopic system, because of which its 
spectrum can be treated as continuous, similarly to the density of states 
of macroscopic statistical systems [53]. Therefore, the summation over 
$k$ in the above formulas should be understood as integration over 
this multi-index. We shall consider two typical distributions, the 
Gaussian and Lorentz ones, and two opposite measurement procedures, 
instantaneous and continuous. This choice is based on the following 
arguments. A measuring device, being a macroscopic object, contains a 
large number of elements acting randomly on the system. As is known from
the central limit theorem, the action of a large number of random elements
is well described by Gaussian distribution. Because of this, the effect 
density of measuring devices is commonly represented by Gaussians [39,54-56].   
Lorentzian distribution arises when measurement is realized by means of
optical beams [40]. Two usually considered types of measurements are 
instantaneous [39,40] and continuous [39,41,55] measurements.

In the case of the Gaussian distribution
\be
\label{56}
 p_{mn}^G(x) = \frac{1}{\sqrt{2\pi}\;\sgm} \; \exp \left (
-\; \frac{x^2}{2\sgm^2} \right ) \;  ,
\ee
we have the decoherence factor
\be
\label{57}
 D_{mn}^G(t) = \exp \left \{ -\; \frac{\sgm^2}{2} \; M^2 \vp^2(t) 
\right \} \;  .
\ee
Here the standard deviation $\sigma$, in general, can depend on the 
indices $m$ and $n$. But for the simplicity of notation, this dependence 
is not shown explicitly. 

While for the Lorentz distribution
\be
\label{58}
 p_{mn}^L(x) = \frac{\sgm}{\pi(x^2+\sgm^2)} \;  ,
\ee
the decoherence factor is
\be
\label{59}
D_{mn}^L(t) = \exp\{ - \sgm M \vp(t) \} \;   .
\ee
  
When each act of the measurement procedure is instantaneous, such that
\be
\label{60}
f_j(t) = \dlt(t-t_j) \;   ,
\ee
then Eq. (39) is the unit-step function
\be
\label{61}
 \vp_j(t) = \Theta(t-t_j) \; .
\ee
Consequently, after the last measurement at time $t_M$, 
\be
\label{62}
 \vp(t) = 1 \qquad ( t > t_M) \;  .
\ee
Then for the decoherence factor (57), in the case of the instantaneous 
measurements, we get
\be
\label{63}
 D_{inst}^G(t) = \exp\left ( -\; \frac{\sgm^2}{2} \; M^2 
\right ) \;  ,
\ee
where the indices $m$ and $n$ are omitted. And for the decoherence 
factor (59), we find
\be
\label{64}
 D_{inst}^L(t) = \exp\left ( -\sgm M \right ) \;   .
\ee

Another situation happens in the opposite case of a single but 
continuous measurement, when
\be
\label{65}
 f_j(t) = 1 \; , \qquad M = 1 \;  ,
\ee
so that
\be
\label{66}
 \vp_j(t) = \vp(t) = t \;  .
\ee
Then the decoherence factor (57) is
\be
\label{67}
 D_{cont}^G(t) = \exp \left \{ -\; \frac{1}{2} \left (
\frac{t}{t_{dec} } \right )^2 \right \} \;  ,
\ee
with the {\it decoherence time}
\be
\label{68}
 t_{dec} \equiv \frac{1}{\sgm} \;  .
\ee
And the decoherence factor (59) becomes
\be
\label{69}
D_{cont}^L(t) = \exp \left ( -\; \frac{t}{t_{dec} }
\right ) \;   ,
\ee
with the decoherence time as in Eq. (68).   

In this way, a finite number of instantaneous measurements or 
continuous measurements during a finite interval of time lead to 
a partial decoherence in expression (55) of observable quantities. 
If the number of instantaneous measurements infinitely increases or 
the time of accomplishing continuous measurements tends to infinity, 
then decoherence becomes complete:
\be
\label{70}
  \lim_{M\ra\infty} \lgl \; \hat A(t) \; \rgl = \lim_{t\ra\infty}
\lgl \; \hat A(t) \; \rgl = \sum_n \rho_{nn} A_{nn} \; .
\ee
Since the limit in Eqs. (70) does not depend on time, this also implies 
equilibration. 

The matrix $\rho_{mn}$, according to Eqs. (22) and (28), generally, 
depends on the initial state of the total composite object, system plus 
device, since
$$
 \rho_{mn} = \sum_k \; \lgl \; mk \; | \; \hat\rho_{AB}(0) \;
| \; nk \; \rgl \; .
$$
The dependence on the measuring device disappears if at the initial 
time the studied quantum system and the measuring device are not 
entangled, so that 
$$
 \hat\rho_{AB}(0) = \hat\rho_A(0) \bigotimes 
\hat\rho_B(0) \;  .
$$ 
Then because of the normalization condition
$$
 \sum_k \; \lgl \; k \;| \; \hat\rho_B(0)\; |\; k \;\rgl = 1 \; ,
$$
the matrix element 
$$
\rho_{mn} = \lgl\; m \;| \; \hat\rho_A(0) \; |\; n \; \rgl 
$$
contains information only on the system initial state. 
 
In the limit of infinite number of instantaneous measurements or large 
time of continuous measurement, the observable quantities reduce to 
the averages involving only the diagonal elements $\rho_{nn}$, which 
demonstrates the importance of such diagonal terms [57]. The values of 
$\rho_{nn}$ can be found by defining the corresponding representative 
ensemble uniquely describing the system [58]. The diagonalization occurs 
not because of imposing some additional averaging over the random phases 
of initial states [59], or because of time-averaging resulting in ergodic
averages [34], but happens naturally as a consequence of repeated 
measurement actions.

\section{Discussion}

The evolution of a quantum system, subject to the action of 
nondestructive measurements, is considered. Two types of measurement 
procedures are analyzed, instantaneous measurements and continuous 
measurements. A finite number of instantaneous measurements or continuous 
measurements during a finite time interval lead to partial decoherence. 
But if instantaneous measurements are repeated infinite number of times 
or a continuous measurement lasts infinite time, then there happens 
complete decoherence and equilibration.

The interaction of a quantum system with a measuring device is principally
different from its interaction with a bath. The latter is usually represented 
by a large equilibrium system. Contrary to this, the measurement procedure 
is a nonequilibrium process. Therefore the interaction part of the total 
Hamiltonian is time-dependent. This does not allow one to write down the 
evolution operator in the standard exponential form. To solve the evolution
equation, we have to invoke the Lappo-Danilevsky theory [49].   

We have considered the evolution of quantum systems under the action of 
{\it nondestructive measurements}. Then the system, starting form an initial
nonequilibrium state, evolves into a state that is partially decoherent. The 
level of decoherence depends on the type and number of nondestructive 
measurements. Real measurements are usually supposed to be nondestructive 
in order not to destroy the studied system. Of course, a measurement, being 
an external perturbation, could also be destructive. Moreover, it is possible 
to study the situation opposite to that treated in the present paper. It is 
admissible to investigate the problem of how an equilibrium system develops
being subject to the action of a time-dependent external perturbation driving
the system out of equilibrium [60]. But such destructive perturbations, 
generally, are not characterized as measurements. And this opposite process 
of system destabilization is a principally different problem.
  
It is worth stressing that absolutely isolated systems do not exist. The notion 
of an absolute isolation is self-contradictory, since the statement that a 
system is isolated implicitly assumes that the fact of the system isolation 
is proved by observations. In the other case, such a statement cannot be true. 
To be true, a statement must be confirmed by observations. But the latter, 
even being accomplished by nondestructive measurements, influence the system, 
leading to its, at least partial, decoherence. The absence of absolutely 
isolated systems and their decoherence, caused by observations, result in the
system equilibration.
 
The fact that the system isolation must be proved by measurements
does not contradict the possibility of treating a system as being
isolated during a finite time duration. For instance, one can prepare a
system in a well defined initial state by a prescribed preparation
method. Then, after a finite time, one accomplishes a measurement by a
nondestructive state tomography and determines the final system state.
Supposing that in between these two measurements, the initial and final
one, the system has not been influenced by any other disturbances, it
could be possible to assume that it has been isolated during this finite
time interval. Such an assumption is nothing but a particular case of
the finite number of measurements, as has been considered above. But, in
any case, one has to accomplish at least two measurement procedures in
order to assume that the system has been isolated for the finite time
between these measurements. In such a case of a finite temporal
observation, the system, as is explained above, does not equilibrate and
exhibits only partial decoherence.

\vskip 5mm

{\bf Acknowledgement}
\vskip 3mm

Financial support from the Russian Foundation for Basic Research is 
appreciated. I acknowledge useful discussions with E.P. Yukalova.

\end{document}